\documentclass[fleqn,10pt]{wlscirep}
\usepackage[utf8]{inputenc}
\usepackage{ulem}
\usepackage[T1]{fontenc}
\title{Speckle pattern analysis of PVK:rGO composite based memristor device  }

\author[1,$\dagger$]{Ramin Jamali}
\author[1,$\dagger$]{Madeh Sajjadi}
\author[1,$\dagger$]{Babak Taherkhani}
\author[1,*]{Davood Abbaszadeh}
\author[1,2,*]{Ali-Reza Moradi}

\affil[1]{Department of Physics, Institute for Advanced Studies in Basic Sciences (IASBS), Zanjan 45137-66731, Iran}
\affil[2]{School of Nano Science, Institute for Research in Fundamental Sciences (IPM), Tehran 19395-5531, Iran}
\affil[*]{moradika@iasbs.ac.ir,~d.abbaszadeh@iasbs.ac.ir}

\affil[$\dagger$]{These authors contributed equally to this work.}

\begin{abstract}
The memristors are expected to be fundamental devices for neuromorphic systems and switching applications. For example, the device made of a sandwiched layer of poly(N-vinylcarbazole) and reduced graphene composite between asymmetric electrodes (ITO/PVK:rGO/Al) exhibits bistable resistive switching behavior. Depending on the resistance state of the (ON-state or OFF-state) at a constant applied voltage, it may show two different resistivities. The performance of the memristor can be optimized by controlling the doping amount of graphene oxide in the PVK polymer. To assess the performance of the device, when it switches between ON and OFF states,  optical characterization approaches are highly promising due to their non-destructive and remote nature.
Here, we characterize the memristor device by the use of speckle pattern (SP) analysis. 
The speckle pattern is the interference of multiple light waves with random relative phases, which is generated via different mechanisms such as scattering from diffusive materials.  
Therefore, SPs can be used to investigate such samples as they include a huge amount of information to be statistically elaborated. 
The experimental paradigm includes \textit{in situ} acquisition of SPs of the PVK:rGO in different states followed by statistical post-processing toward examining its conduction mechanism.
We express the processing results in terms of several statistical parameters. The variations in these parameters are attributed to the resistance state of the PVK:rGO samples under the applied voltage with regard to the physical switching mechanism of the device. The resistance/conduction state, in turn, depends on the activity and properties of PVK:rGO memristors as well as the additional non-uniformities induced through the variations of density of carriers.
The present optical methodology can be potentially served as a bench-top device for characterization purposes of similar devices while they are operating. 
\end{abstract}
\begin{document}

\flushbottom
\maketitle
%
%
\thispagestyle{empty}

\section{Introduction}
Rapid developments in the  storage technology requires significant focus on addressing the  issues of the traditional  semiconductor industries, such as processing cost and device printability \cite{forrest2004path,buga2021review}. 
Organic or polymeric materials are preferred  to inorganic materials due to their good scalability, simple processing, low cost, low-power operation, flexibility, large storage capacity, 3D stacking capability, and easy processing \cite{raymo2002digital,reed1997conductance,yang2006electrical,yang2004organic,service2001organic,scott2004there,park2007blue}.

An electronic memory device that stores information as binary numbers, is a type of semiconductor storage space that has a fast response and compact size, which can be read and written when connected to a central processing unit (CPU). 
In silicon-based electronic memory, information is stored based on the amount of the charge stored in the memory cells. 
\textcolor{black}{Newly emerged}    electronic memories store data in a completely different way, e.g., based on different electrical \textcolor{black}{resistivity/}conductivity states \textcolor{black}{demonstrated as} ON and OFF \textcolor{black}{states} in response to an applied electric field \cite{zhang2015organic}.
Regarding the aspiration for new data storage technologies, ferroelectric random access memory (FeRAM) \cite{setter2006ferroelectric}, magnetoresistive random access memory (MRAM)  \cite{de2002technology}, phase change memory (PCM) \cite{hudgens2004overview}, and different kind of organic memories have appeared on the scene of the information technology industry \cite{moller2003polymer}.
The novel technologies are based on the electrical bistability of materials arising from changes in certain intrinsic properties, such as magnetism, polarity, phase, reduction-oxidation, conformation, and conductivity in response to the applied electric field. 
{\color{black}Controlling these memory properties is achieved by tuning the physical/chemical properties of materials. In other words, two stable states, physically or chemically,  can be achieved upon application of  \textcolor{black}{magnetic or} electric fields, which  is the basis for the fabrication of the memory devices.}
All types of memory characteristics, from volatile memory properties, such as dynamic random access memory (DRAM) and static random access memory (SRAM), to non-volatile memory properties, such as flash memory and  write-once-read-many-times (WORM) types, have been reproduced in resistive memory systems \cite{zhang2015organic}. 
Devices incorporating switchable resistive materials are generally classified as resistor-type memory and resistive random-access memory (RRAM). 
Unlike the conventional memory technologies with the memory effects associated with a specific cell structure, {\color{black} resistive memories, so-called  memory-resistors or  memristors, store data totally in a different form which is based on the change of resistance.} 
\textcolor{black}{Despite the capacitor-type and transistor-type polymer memories, the RRAMs do not need to be  integrated into basic CMOS logic circuits \cite{ling2008polymer}.}

Most of the reported materials used for the fabrication of memristors  are organic molecules, polymeric materials such as polyimides, conjugated and non-conjugated polymers, fullerenes and graphene  composited polymers, and hybrid organic/inorganic compounds like  polymers with metal nano-particles \cite{zhang2015organic}. \textcolor{black}{Nowadays, investigation of the devices made of the aforementioned materials is among the hot research topics, as such devices are the basis for neuromorphic systems.} 

{\color{black} In this paper, we report the fabrication and study of a memristor that is fabricated using the composite of  PVK and graphene oxide. We investigate its electron transport through measurement of current-voltage (J-V) characteristics which is a simple and straightforward way to analyze an electronic device's performance. J-V measurement provides the basic information about the memristor's memory action under different voltage and current conditions. Basic information like switching voltage, memory window, and their repeatability are important memristor characteristics that can be captured through repetitive I-V scans. To further in-depth investigation of our memristors, we study them under OFF and ON states, for which the physical properties of the materials like dielectric or electron transport, which, in turn, changes upon the change of the resistance are used.}

For the investigation of the fabricated memristor, we introduce dynamic speckle pattern (SP) analysis as a remote and nondestructive optics-based method for resistive switching characterization of PVK:rGO memristor devices.
The SP includes a random distribution of high contrast and fine-scale bright intensity grains surrounded by a network of dark regions. SP may be formed upon the interference of numerous coherent light waves that are out of phase with each other. This type of interference can be generated through several processes, such as the scattering of laser light off a rough surface, transmission of laser light through a diffusive or translucent matter, and superposition of propagating spatial modes out of a multi-mode fiber \cite{goodman2007speckle}. Therefore, SP delves into the intricacies of two specific types: transitive and reflective \cite{goodman2007speckle,braga2003assessment}.

In the first glance, a SP  appears just as a noise in optical systems to be removed by performing a statistical averaging. Indeed, in some optical imaging systems the inherent fluctuations of laser light or its scattering from a set of contaminations within the optical train case disturbing SPs. In order to achieve a clean and  high contrast image in such systems the noises are reduced by reducing the coherency of the source through using a rotating diffuser or  by subtracting the noises in Fourier space. 
Even though these patterns appear random, the analysis of SPs can reveal intriguing applications. Techniques based on SPs have been recognized as flexible tools for the purposeful investigation of a wide range of physical, chemical, and biological phenomena\cite{aizu1991bio,fujisawa2009temperature}.
When it comes to SPs produced by the scattering of laser light off a rough surface, each point within the illuminated area serves as a source of secondary waves and carries information about the surface. Consequently, the  recording device captures comprehensive information about the surface. The study of these SPs provides valuable insights into the properties of the light source, the medium, and the imaging system. 
Based on the configuration for SP acquisition and the associated analysis three different approaches based on  SP analysis are developed: speckle imaging \cite{loutfi2020real,li2021transmissive}, dynamic speckle  pattern  analysis  \cite{jamali2023surface,pedram2023evaluation,abbasian2024dynamic}, and secondary speckle analysis \cite{kalyuzhner2022remote,duadi2020non}. 
 
In this research, we employ dynamic SP  analysis to examine the conduction mechanism  in the memristor devices during their operation. 
Dynamic SPs emerge when a material under laser illumination exhibits any form of activity. Understanding the origins and characteristics of dynamic speckles can enhance our knowledge about the internal dynamics of the phenomena. The more we know about the internal dynamics of the samples, the better our insights will be in controlled experiments and simulations to evaluate how these dynamics are reflected in the evolution of the speckle \cite{arizaga1999speckle}.
The formation of the SP is achieved through two approaches: subjective and objective. The subjective SP is observed when a coherent light beam illuminates a rough surface, resulting in the SP on the ``image plane''. The detailed structure of this pattern depends on the parameters of the viewing system. On the other hand, the objective SP occurs when a laser light beam scatters a light field without a lens, and a rough surface falls on another surface.  Given the aforementioned comprehensive inclusion of information in the recorded dynamic SPs, the SP-based techniques have proven to be versatile tools for studying various chemical, biological, and physical phenomena \cite{goodman1976some,mohan2009n}. 
Detection, monitoring and investigating of electric current, blood flow, seed health, scaffold activity, characterization of nanocomposites, alignment of multicomponent lipid, fruit ripeness, pitting corrosion, paper crumpling, evaluating tissue viscoelastic properties, adhesive drying, and parasite motility, among others, are some of these interesting applications  \cite{nazari2023laser,fujii1985blood,rabal2018dynamic,rad2020non,jamali2023surface,panahi2022detection,romero2009bio,pedram2023evaluation,rad2019speckle,hajjarian2012evaluating,ansari2016following,pomarico2004speckle}. Another fascinating application of SPs, beyond their use to extract information about their generating mechanism or sample, is the use of SPs, disregarding the mechanism that they are generated, for collective manipulation of microscopic objects \cite{volpe2014speckle,jamali2021speckle,sadri2024sorting}. Dynamic speckle pattern analysis, specifically,  has been  applied  in electrical engineering. Cyclic voltammetry, low power near-sensor, controllable modulation, and realizing spike-timing are some of the applications in this field  \cite{volkov2016cyclic,smagulova2018low,xie2023controllable,suresh2019realizing}.
With respect to common  characterization techniques in electrical engineering and micro-electronics, SP-based methods  offer several advantages,  such as real-time data collection, non-destructive and non-contact nature, inclusion of  integrated spatiotemporal information.
Moreover, this technique does not have phototoxic effects on the sample as it uses a very low laser power for illumination. A remarkable feature of dynamic speckle analysis is that it can be used for extended studies,  lasting even for  several days, provided that  the environmental conditions for the different types of specimens are maintained and the laser power stability is secured. Following this line, here, we introduce the method to characterize  memristors. 
{\color{black} Our samples are made of active materials and two electrodes. The active materials are sandwiched between a transparent bottom electrode of Indium Tin Oxide (ITO) glass and semitransparent Aluminum top electrode. The active layer of the samples is a composite of Poly(N-vinylcarbazole) (PVK) and reduced Graphene Oxide (rGO), which is spin-coated onto the ITO glass. This active layer facilitates the flow of electrons between the electrodes with different resistivity states. Finally, a layer of Aluminum  is evaporated through a shadow mask as the top electrode on top of the active layer in a vacuum chamber. }
To study the effect of steam on the memristor, we  examine  various statistical parameters , including Time History Speckle Pattern (THSP), Co-occurrence Matrix (COM), Inertia Moment (IM), roughness parameters such as Kurtosis and Skewness, Absolute Value of the Differences (AVD), and Contrast. 

In  Section \ref{matmeth},   Materials and Methods, we provide details of the sample preparation, experimental setup and the numerical statistical processing of SPs. There,  the theoretical overview of the statistical analysis of dynamic SPs is presented and the aforementioned parameters are systematically defined and discussed. 
Section \ref{results} showcases the experimental findings and discusses the results of the analysis. The paper is wrapped up with a Conclusion in Section \ref{conc}.


\section{Materials and Methods}
\label{matmeth}
\subsection{Preparation of the PVK-graphene composites}
Graphite oxide is prepared by the Hummers method from graphite and is dried for a week over phosphorus pentoxide in a vacuum desiccator before use \cite{hummers1958preparation}. Dried graphite oxide (100 mg) is suspended in anhydrous DMF (10 ml), treated with phenyl isocyanate (4 mmol) for 1 week while being stirred at room temperature, and recovered by filtration through a sintered glass funnel. Stable dispersion of the resulting phenyl isocyanate-treated graphite oxide materials is  prepared by ultrasonication in DMF. Poly(N-vinyl carbazole) (PVK, purchased from Aldrich, Mw = 1,100,000) is solved in Chlorobenzene and added to these dispersions and is dissolved with stirring. A highly homogeneous black dispersion of PVK:rGO composites are obtained after reduction and finally achieved by the addition of hydrazine (0.1 ml in 10 ml of DMF) at 80 $^{\circ}$C for 24 h.

\subsection{Device fabrication and characterization}
Indium-tin oxide (ITO)-coated glass substrates of 2$\times$2 cm in size are precleaned by ultrasonication for 15 min each with deionized water, ethanol, and isopropanol, in that order. The above-mentioned  DMF solution of PVK:rGO is spin-coated onto the ITO glass, followed by solvent removal in a vacuum oven at 0.6 Torr at 80 $^{\circ}$C for one day. Aluminum (thickness $\approx$ 200 nm) is thermally evaporated onto the film surface at about 10$^{-5}$ Torr through a shadow mask to form 0.9$\times$0.9, 0.7$\times$0.7, 0.5$\times$0.5 and 0.3$\times$0.3 mm$^2$ top electrodes, as shown in Fig. \ref{Figure01}(b). All electrical measurements are carried out on devices with the 0.9$\times$0.9 mm$^2$ top electrode, using a RADstat 1000 potentiostat/galvanostat under ambient conditions.

\subsection{Experimental setup}
\begin{figure*}[t!]
	\begin{center}
		\includegraphics[width=0.7\linewidth]{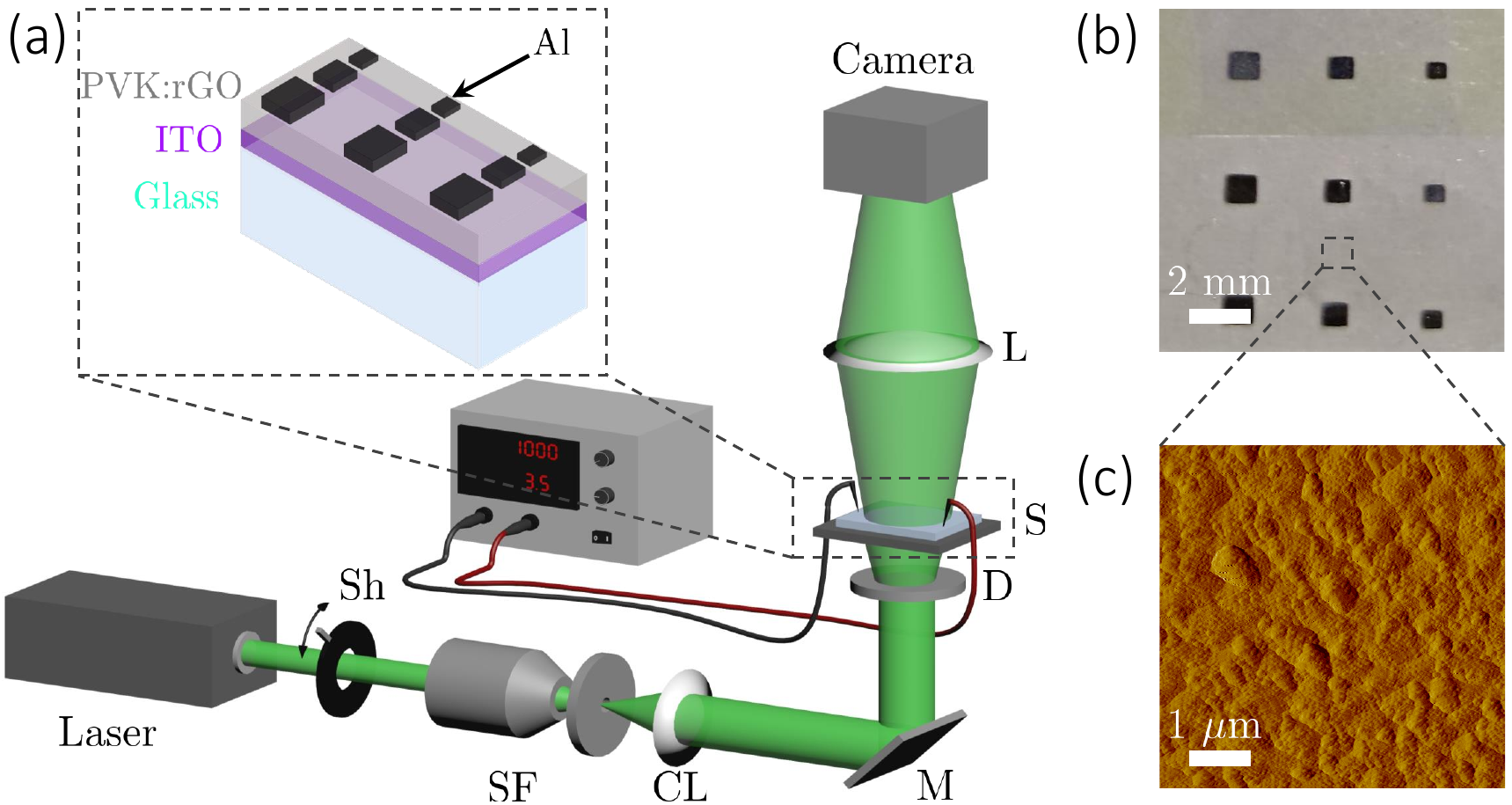}
		\caption{(a) Schematic of the experimental setup  for dynamic speckle pattern analysis in forward scattering mode; Sh: shutter; SF: spatial filter, CL: collimating lens, M: mirror, D: diffuser, S: sample,  L: collecting lens, and PS: power supply. Inset: Enlarged view of the sample structure; Al: aluminum, PVK:rGO: polyvinylcarbazole-reduced graphene oxide, ITO: indium tin oxide. (b) Top view picture of the actual PVK‌ graphene memristor sample. (c)  The 2D micrograph of atomic force microscope (AFM) of PVK:rGO from  the selected area shown in panel (b).}
		\label{Figure01}
	\end{center}
\end{figure*}
For SP analysis based assessment of most of  specimens it is possible to consider either backward or forward scattering light. The forward setup exhibits less sensitivity compared to the backward scattering setup \cite{rabal2018dynamic}, and its use is restricted to instances where the specimens are sufficiently transparent to allow light transmission. Indeed, SP analysis examination remains consistent across both setups, however, the proper choice depends also on the specific sample or application. In this research, according to the above points and given the sample specifications we employ forward scattering arrangement. Figure \ref{Figure01} (a) shows the schematic of the optical setup. 
The laser beam (DPSS Lasers, 532 nm, 100 mW) passes through the spatial filter (SF), collimated with the collimating lens CL (focal length=75 mm), and is redirected onto a diffuser (D) by the use of 45$^{\circ}$ mounted mirror M. The beam  subsequently illuminates the sample (S) after passing through the diffuser. Light scattered from the sample generates the speckle field which is collected by the collecting lens (L, focal length=100 mm) and  captured on a digital camera (Thorlabs, DCC3240M,  1.3 Megapixels, 8-bit dynamic range, 5.3 $\mu$m square pixel pitch). 
The digital camera is configured to record a 700$\times$700 pixel area at its center with an exposure time of 0.20 ms. The  laser source possesses adequate coherence and stability throughout the experimental procedure. An important requirement for a successful SP based analysis is the intensity stability of the illumination source. In addition to the inherent stability feature of  the used laser source,  we turn the laser on  at least one hour prior to the experiment to secure the required intensity stability.  Then, during the experiments the laser remains operational and when needed, instead, the laser beam is obstructed using the shutter Sh. 
The beam uniformity is verified by positioning a mirror at the sample stage and gathering the back-reflected light with the camera for approximately few minutes. 

In the inset of Fig. \ref{Figure01}(a) a magnified scheme of the structure of  PVK:rGO memristor sample is shown.
Figure \ref{Figure01}(b) is a top view picture of the actual  sample, and  Fig. \ref{Figure01}(c) is  an atomic force microscope (AFM) micrograph image   of PVK:rGO from the selected area shown in panel (b). 

\subsection{Numerical processing}
Observing dynamic SPs reveals their spatiotemporal fluctuations, which is a consequence of the materials' structure and activity. The interaction of coherent light with materials can result in absorption, reflection, and scattering, and these events are the primary drivers of SP formation. Considering the sufficient stability of the laser's intensity and wavelength the activity detected by dynamic SPs in materials is attributed to their inherent activities.
Therefore, analysis of  sequences of the dynamic SPs yields valuable insights and information about the samples. 
The activity of the sample, particularly in bio-materials, can be unveiled via various analytical outcomes. Dynamic activity is triggered when the sample alters its properties due to local fine structure movements, configuration changes, movement of the scattering centers, refractive index variations, inherent motility, and so on. 
This study is designed to elucidate the temporal activity of PVK:rGO memristors.
 To achieve this, the captured SPs are converted to successive image formats and are subjected to  numerical processing. The results will lead to  computation of several  statistical parameters which are outlined in the following. These parameters are then used to assess the samples for their characterization, categorization, defect recognition, etc. 
\subsubsection{Time history of speckle patterns (THSP)}
THSP is a 2D matrix that encapsulates the temporal progression of consecutive SPs. THSP is formed by random selection of M points in each pattern of the N successively acquired SPs, and arranging such M $\times$ 1 columns next to each other. Therefore,  the new composite image, THSP,  will have  M $\times$ N size; the rows (M) of this matrix symbolize the set of points, and the columns (N) represent their intensity state at each sampled moment, hence signifies the temporal progression and the total number of patterns. THSP is utilized to characterize the activity in samples, with activity manifesting as intensity changes in the horizontal direction. Thus, greater intensity variations in the THSP horizontal lines correspond to samples with higher activities \cite{braga2008time}.
\subsubsection{Co-occurrence matrix (COM)}
Besides the graphical activity visualization that THSP provides, more importantly, it is employed for additional numerical outcomes, such as the co-occurrence matrix (COM), inertia moment (IM), absolute value of the differences (AVD),  and so on.  
The COM represents the probability of transitioning between intensity values in two adjacent pixels of THSP. This transition histogram of intensities illustrates the sample's activity as the dispersion of non-zero values beyond the main diagonal.
The COM  is defined as follows: 
\begin{equation}
	{\rm{COM}}(i,j)=\sum_{m=1}^{M}\sum_{n=1}^{N-1} 
	\begin{cases}
		1, & \text{if ~~ {\rm{THSP}}$(m,n) = i$} \\
		& \text{and~~{\rm{THSP}}$(m,n+1) = j$,}\\
		0, & \text{otherwise.}
	\end{cases}
	\label{eq:1}
\end{equation}
Here, $i$ and  $j$ denote the intensity of two adjacent pixels.
For an active sample, the intensity values evolve over time, the non-zero elements near the main diagonal increase, and the matrix takes on a cloud-like appearance. Conversely, for a low-activity sample, the matrix values are concentrated around the main diagonal \cite{zdunek2014biospeckle,arizaga1999speckle}.
\subsubsection{Inertia moment (IM)}
The IM quantifies the dispersion of values around the  COM principal diagonal. IM is defined as the accumulated COM matrix values that are multiplied by the square distance to the original diagonal and normalized \cite{zdunek2014biospeckle}:
\begin{equation}
	{\rm{IM}} = \sum_{i}^{}\sum_{j}^{} \frac{{\rm{COM}}(i,j)}{\sum_{m}{\rm{COM}}(i,m)}|i-j|^2.
	\label{eq:2}
\end{equation}
The normalization process minimizes the impact of heterogeneity in the analyzed images and ensures that the sum of the occurrence values in each line of the COM equals  to 1 \cite{arizaga1999speckle}. The IM measurement serves as a valuable statistical tool for estimating the sample activity, and  increasing IM values corresponds to increased sample activity.  
\subsubsection{Absolute value of the differences (AVD)}
The AVD is a first-order statistical moment, which is an adaptation of the IM process. The AVD process quantifies a measurement relative to the absolute intensity leap value between successive images, specifically between consecutive images  I$_k$ and I$_{k+1}$. It operates on the principle that the accumulation of differences is the primary information sought, and  is defined as  \cite{ansari2013assessment,braga2011evaluation}:
\begin{equation}
	\rm{AVD} = \sum_{i}^{}\sum_{j}^{} \frac{{\rm{COM}}(i,j)}{\sum_{m}{\rm{COM}}(i,m)}|i-j|.
	\label{eq:3}
\end{equation}
Indeed, Eq. \ref{eq:3} demonstrates that the value of AVD is a weighted measure of the absolute difference between the intensities of two different pixels. Taking the absolute  highlights the principle that the AVD method is fundamentally concerned with the magnitude of intensity changes between successive images.
\subsubsection{Contrast}
Additional to the aforementioned parameters that are extracted from the computation of the THSP matrix, several other statistical parameters may be also used to assess the SPs, as these patterns essentially are 2D statistical intensity  distributions. 
Contrast is employed to identify the movement activity within a sample. As such, the contrast of intensity values over time is utilized to construct activity images. The  SPs contrast image is defined as follows \cite{moreira2014quality}:
\begin{equation}
	{\rm{Contrast}} = \frac{\sigma_{x,y}}{\langle I \rangle},
	\label{eq:4}
\end{equation}
where,  $\langle I \rangle$ and $\sigma$ denote the temporal mean value and standard deviation of the SP matrices, respectively. 
\subsubsection{Roughness} 
Considering the SP intensity matrix as a 2D statistical distribution, roughness parameters may also be defined and used for SP-based evaluation purposes, independently of THSP.   It is remarkable that  in the case of speckle formation via scattering of laser light from a rough surface, the SP matrix roughness is correlated to the roughness of the sample \cite{jeyapoovan2012statistical}. This correlation between  intensities and surface roughness provides a powerful tool for surface characterization. However, the roughness of the SPs can also separately measure  the internal activities of the samples, regardless of the process that led to the formation of SPs  \cite{jamali2023surface,pedram2023evaluation}.  
The set of Average roughness (R$_{\rm{av}}$),  Root mean square (R$_{\rm{rms}}$),  Skewness (R$_{\rm{sk}}$), and Kurtosis (R$_{\rm{ku}}$) parameters provide a comprehensive and reliable  insight to the associated distribution. These parameters are the first, second, third, and fourth  moments of the data deviation  to the average value of the distribution, respectively.  

The average roughness (R$_{\rm{av}}$) represents the average deviation of intensities from the mean value across all data points:
\begin{equation}
	{\rm{R_{\rm{av}}}} = \frac{1}{P~Q} \sum_{p=1}^{P} \sum_{q=1}^{Q}|I(p,q) - \langle{I}(p,q)\rangle|.
	\label{eq:5}
\end{equation}
This parameter provides a measure of the average variation in pixel intensity in the SPs. On the other hand, the root mean square  of roughness (R$_{\rm{rms}}$) signifies the standard deviation of the distribution:
\begin{equation}
	{\rm{R_{\rm{rms}}}} = \bigg[\frac{1}{P~Q} \sum_{p=1}^{P} \sum_{q=1}^{Q}\left[I(p,q) - \langle{I}(p,q)\rangle\right]^2\bigg]^{\frac{1}{2}}, 
	\label{eq:6}
\end{equation}
where $P$ and $Q$ denote the horizontal and vertical dimensions of the SPs, $p$ and $q$ are the counts of pixel numbers, and $I$ signifies the intensity across the SPs. This parameter provides a measure of the spread of the pixel intensity values around the mean, offering insights into the variability of surface roughness. The average roughness  and root mean square of roughness represent  the standard deviation of the distribution. 
R$_{\rm{sk}}$ and R$_{\rm{ku}}$ are the two additional metrics that  offer valuable descriptive information about the intensity distribution: 
\begin{equation}
	{\rm{R_{\rm{sk}}}} = \frac{1}{P~Q~S_2^3} \sum_{p=1}^{P} \sum_{q=1}^{Q}\left[I(p,q) - \langle{I}(p,q)\rangle\right]^3,
	\label{eq:7}
\end{equation}
\begin{equation}
	{\rm{R_{\rm{ku}}}} = \frac{1}{P~Q~S_2^4} \sum_{p=1}^{P} \sum_{q=1}^{Q}\left[I(p,q) - \langle{I}(p,q)\rangle\right]^4.
	\label{eq:8}
\end{equation}
R$_{\rm{sk}}$ quantifies the symmetry level of the intensity distribution. The skewness of a symmetric distribution is zero as the deviations from average in lower intensities neutralize the deviations in higher intensities. A negative skewness indicates a predominance of valleys, whereas a positive skewness signifies a distribution with high intensities. 
R$_{\rm{ku}}$ gauges the acuteness of the distribution across the pattern. For a standard distribution of intensities, kurtosis is 3. An intensity distribution with pronounced peaks or valleys has a kurtosis exceeding 3, and a wider distribution associated with SPs exhibiting gradual variation has a kurtosis less than 3 \cite{jeyapoovan2012statistical,gadelmawla2002roughness}.


%
 
\begin{figure*}[t]
	\begin{center}
		\includegraphics[width=.7\linewidth]{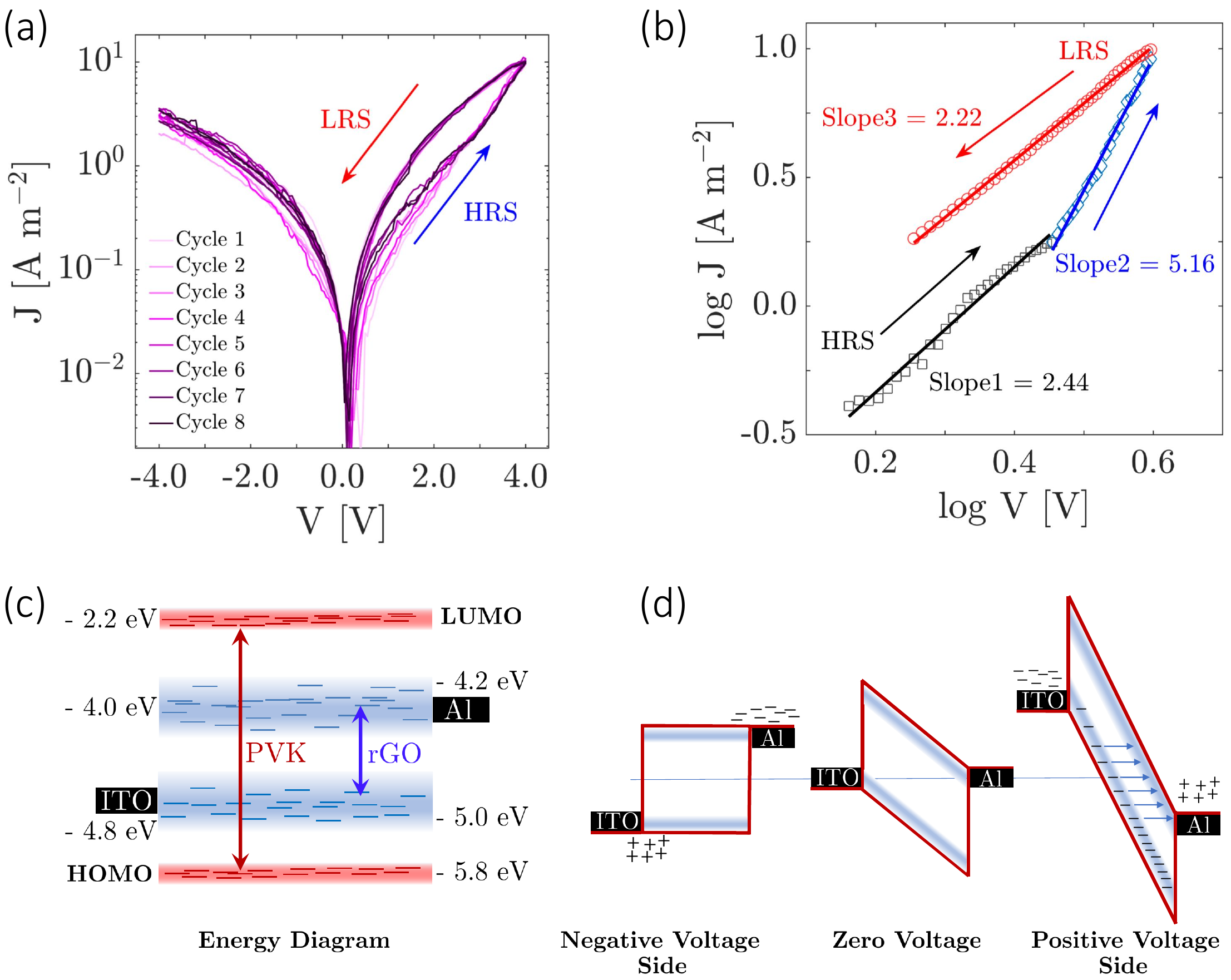}
		\caption{(a) The current density, J, versus the applied voltage, V, for a pristine memristor device as a characteristic of the device with ITO/PVK:rGO/Al structure for 8 different switching profiles for the transition from HRS to LRS. Switching loop of the memristor is obtained without a forming step. (b) Fitted curves on $\log$ J versus V for two different sweeps. Two different slopes are observed for LR and HR states;  the slope 2.25 represents space charge limited current (SCLC) behavior, and the slope 5.05 is the representative for tunneling current. (c) Energy level of the device structure. (d) Switching mechanism according to the J-V plot. The arrows show tunneling of electrons that become substantial above 2.6 V.}
		\label{Figure02}
	\end{center}
\end{figure*}

\section{Results and Discussion}
\label{results}
Figures \ref{Figure02} (a) and (b) show the performance of PVK:rGO memristor. It is deduced from  Fig.  \ref{Figure02}(a) that the memory effect occurs in the positive applied voltages. Moreover, it shows that  the direction of J-V characteristics in the sweep of 0 to 3.5 and 0-volt is counter-clockwise. The device switches from a high resistance state (low-conductivity state or OFF state) to a low resistance state  (high-conductivity state or ON state) at about 2.6 V. Then, the "write" operation is performed on the memory and  by decreasing the applied voltage from 3.5 V the device stores the ON state. 
By sweeping voltage to the negative voltages and back to zero voltage, writing is erased and the device goes back to the OFF state. In Fig. \ref{Figure02} (a) 8 cycles of this procedure, which acts as the mechanism of writing and reading in the memory, are shown. The OFF state of the device can be reprogrammed to the ON state in the subsequent positive voltage sweep, thus completing the write-read-erase-read-rewrite cycle for a non-volatile rewritable memory devices.  

In order to distinguish the aforementioned mechanism in Fig. \ref{Figure02} (b) V-I data is plotted in log-log scale. Figure \ref{Figure02}(b) shows that with increasing the voltage current increases linearly that is the sign of space charge limited current (SCLC) and upon making better or thicker devices this part can be diminished considerably.  As the voltage increases the current increases until 2.6 V with the slope of about 2.44 (high resistance state -HR- or OFF state).  For higher than 2.6 V voltages the current increases with steeper slope (about 5.16) and writing process happens. Subsequently, in the return, with decreasing the voltage the current decreases and the memory remains in the low resistance  (LR) or ON state. The write state can be erased with application of negative voltage or sweeping to the negative voltages. The writing and erasing procedure can be repeated many times for these devices. The 8 cycles of the procedure shown in   Fig. \ref{Figure02} (a)   confirms the fabricated device's reliability and repeatability. Slopes higher than 2 can be due to the trapping effect or the tunneling of carriers. Here, trapping cannot be the case since the trapping makes clockwise hysteresis with no repetition of memory window. Therefore, according the observed memory window and the slope analysis, transport regime in the ON state is the tunneling type.

\begin{figure*}[t!]
	\begin{center}
		\includegraphics[width=\linewidth]{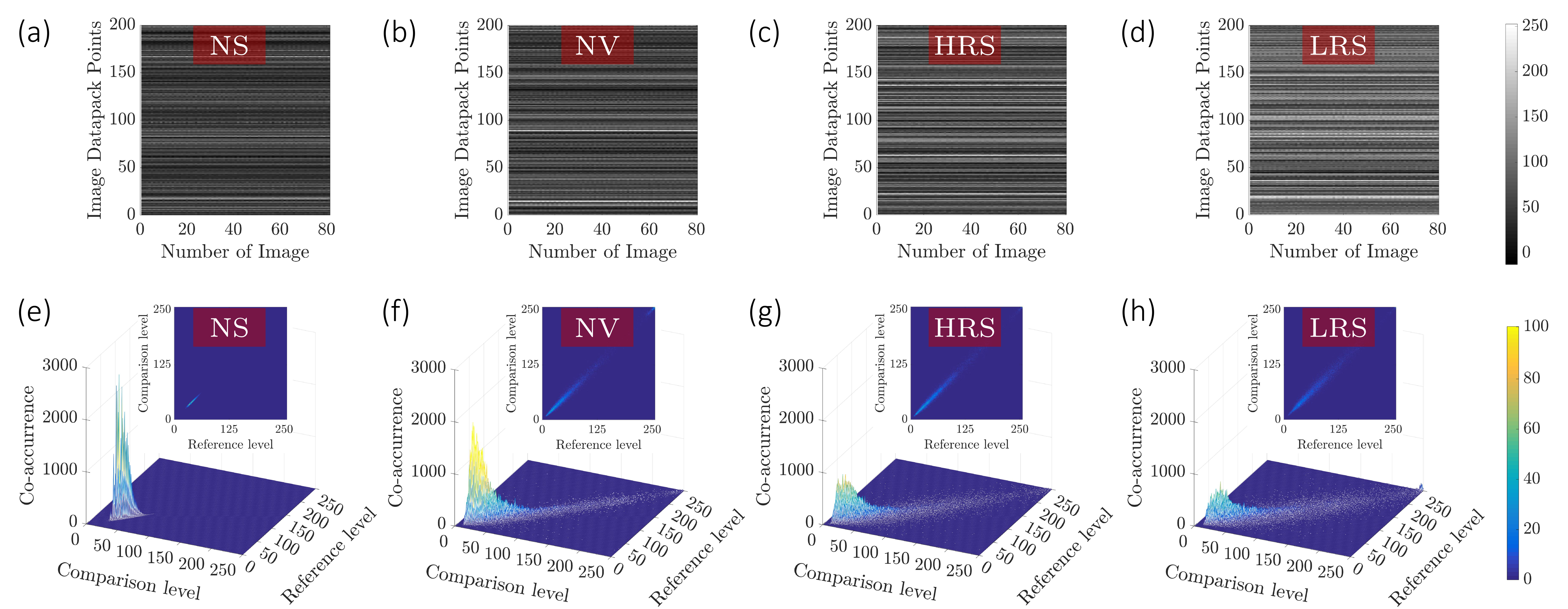}
		\caption{The time history speckle pattern (THSP) and Co-occurrence (COM) matrices of the samples. THSP is formed by tracking 200 random points throughout a collection of 80 SPs of each sample. THSP of (a)
Control experiment when no sample is place in the setup; (b) No voltage (NV) is applied to the fabricated sample; (c) Increasing voltage of cycles or high resistant state  (HRS); (d) Decreasing voltage of cycles or low resistance state (LRS). The associated 2D map and 3D demonstration of COM matrices of (e)
Control experiment when no sample is place in the setup; (f) No voltage (NV) is applied to the fabricated sample; (g) Increasing voltage of cycles or high resistant state  (HRS); (h) Decreasing voltage of cycles or low resistance state (LRS). The comparison level and the reference level demonstrate the intensity levels of $i$ and $j$ in Eq. \ref{eq:1}, respectively.}
		\label{Figure03}
	\end{center}
\end{figure*}

Figure \ref{Figure02}(c) shows the diagram of the energy level structure of the device. As it is shown, PVK is a wide band gap semiconductor, which plays the role of the host for low band gap rGO that is the guest.  In this system transport will occur through host molecules, if the rGO percentage is low, and rGO will play the role of trapping centers for the electrons.  This is due to the sparsely distribution of rGO flakes in the PVK matrix which enhances the energy cost for electrons hopping. When the amount of rGO moieties in the PVK matrix  is increased, their distances will decrease and hopping of the electrons become more feasible. 
 The transition from host to guest in organic semiconductors is  about 0.1\% to 5\%  for guest material  \cite{yimer2008charge,coehoorn2012effects}. 
 According to our finding and the predated studies  \cite{yimer2008charge}, this transition occur around 3\% of rGO in the PVK matrix, which is in agreement to the mentioned findings. The resistive switching and memory effects of PVK:rGO device can be observed only for 3 wt\% rGO in PVK matrix as shown in the (J-V) measurement. This means that, we only see memory effect  in the intermediate transition regime from trapping transport to host-guest transport.
 In other words, it is possible to observe the memory effect only when the transport of electron  happens through both host and guest molecules. Below the 3\% guest, transport purely is based on host and guests trap carriers effectively.
 At guest concentrations of  higher than 3\%, transport of electrons occur purely by guest. To recap the above statement,  doping of PVK with about 1\% rGO decreases the conductivity due to the formation of traps. Further doping (higher than 2-3\%), however, increases the conductivity via guest molecules and we do not observe any memory effect. In other words, at higher doping levels the hysteresis disappear and the device becomes highly conductive. So we can conclude that the memory effect hysteresis just occurs in the intermediate regime. 

We demonstrated the switching mechanism graphically in Fig. \ref{Figure02}(d), in which   the device is considered  under negative (backward), zero and positive (forward) bias. At negative bias, the contact might be ohmic-like. Therefore, the device demonstrate SCLC current. In the forward bias,  electrons and holes experience a barrier causing the current to flow via tunneling, demonstrated by arrows. It is continued  until the device switches on and reaches to its new SCLC limit. The charges that accumulate in the interface of the PVK:rGO/Al facilitate tunneling when the voltage goes higher and keeps the ON state until reverse bias removes them from the interface.

\begin{figure*}[t!]
	\begin{center}
		\includegraphics[width=\linewidth]{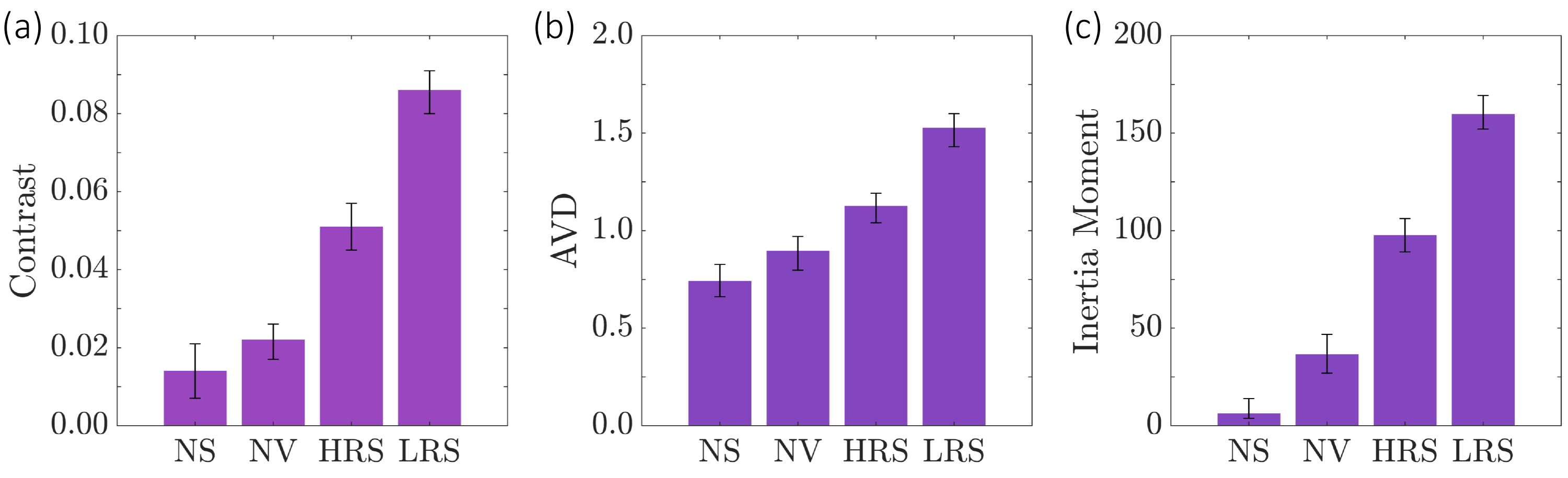}
		\caption{Quantitative SP analysis parameters. (a) Average inertia moment; (b) Average  absolute value of the differences (AVD); (c) Average speckle contrast  for the  four cases of NS, NV, HRS and LRS of PVK graphene memristors. }
		\label{Figure04}
	\end{center}
\end{figure*}

\begin{figure*}[t!]
	\begin{center}
		\includegraphics[width=\linewidth]{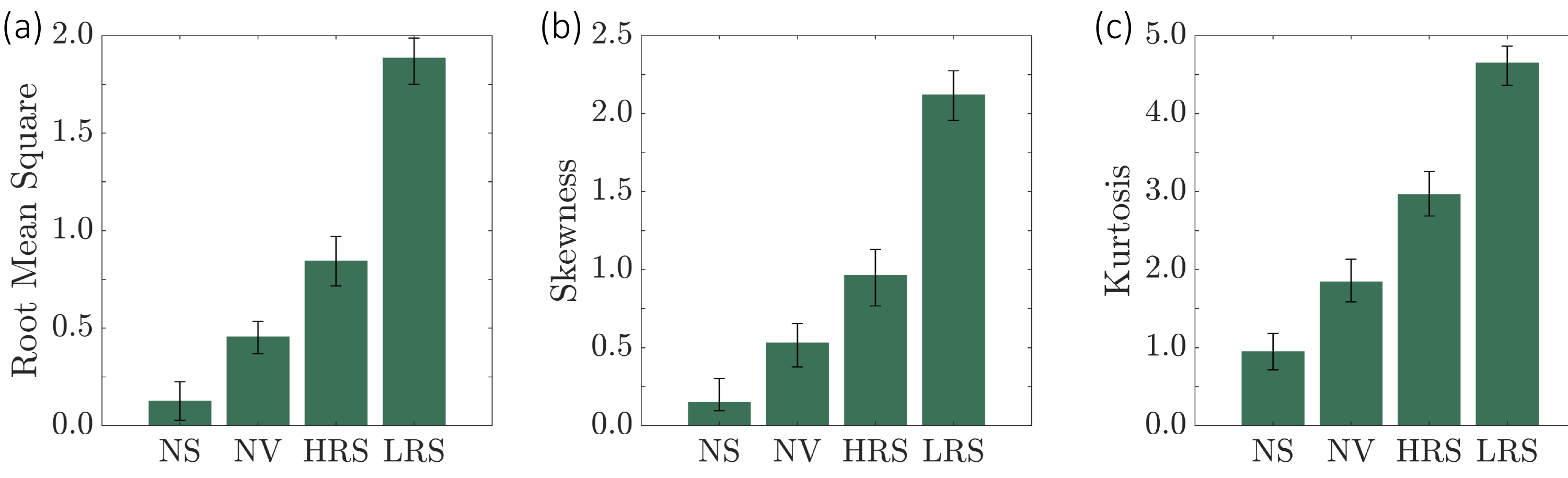}
		\caption{Roughness  parameters of the SP intensity statistical distributions. (a) Average root mean square (RMS); (b) Average  skewness; (c) Average kurtosis  for the  four cases of NS, NV, HRS and LRS of PVK graphene memristors. }
		\label{Figure05}
	\end{center}
\end{figure*}

Characterization  and the explanation of the mechanism using micro-electronic methods provide precise and quantitative information about the performance of the devices. However, the micro-electronic approaches are rigorous, costly, require probe and contact, and in some cases can be destructive. In this paper, we introduce the optical approach of SP analysis as an elegant alternative for assessment of \textcolor{black}{electronic} memristor devices. Despite common micro-electronic approaches, SP analysis is remote, contactless, non-invasive and easy-to-implement.  As explained in Section 2.4 the SPs include a huge amount of statistical information about the samples from which the laser light is scattered and forms the SPs. The analysis of the SPs, therefore, is carried out via statistical process and leads to several statistical parameters that can be used for various comparative studies of the samples. 
We apply the SP analysis on three types of PVK:rGO memristors:  when no voltage is applied (NV), high resistance state (HRS),  low resistance state  (LRS),  and a control case in which no sample is placed (NS). In HRS and LRS the current density is low and high, respectively. 
The SP results are outlined in Figs. \ref{Figure03}-\ref{Figure05}. 
Figures \ref{Figure03} (a-b) show the THSP matrices of the above four samples.  The THSPs are formed by tracking 200 random points throughout a collection of 80 SPs of each sample. The results show that increasing the voltage of cycles, i.e., in high resistances the influence of the sample on the intrinsic intensity fluctuations of the laser light causes the formed SPs to alter  in time strongly. This is deduced from the non-steady intensity lines in the THSP patterns of the cases when a voltage is applied, and the effect is more pronounced  in lower resistances. This visual difference is even more visible with the COM matrices that are shown in Figs. \ref{Figure03}(e-h). Therefore, it is obvious that  with SP analysis, even visually, it is convenient that the different groups of the fabricated memristors may be distinguished or categorized.  The distinction can be performed in a more quantitative fashion if further associated SP analysis parameters are calculated. In Figs.  \ref{Figure04}(a-c) we calculated and plotted three main statistical parameters of IM, AVD and contrast. The parameters are calculated for at least five measurements, and their average along with their standard deviation are reported. Our results show with decisiveness  that the  formation of SPs in the different types of PVK:rGO memristors is highly influenced by the types of the samples. 

We may attribute the  differences between the aforementioned cases to the interaction of the density of carriers with the external light field \cite{falco2010density}. \textcolor{black}{Due to the changes in the density of carriers in the samples, their responses to the incoming laser light and, hence, the SP that they form will be different.} Therefore, it will lead to different associated statistical parameters. 
 
In order to verify the hypothesis for our samples we measure the roughness of the SPs which are shown in Fig. \ref{Figure05}. Second, third, and fourth moments of the deviation from the mean value, i.e., the RMS, the Skewness, and the Kurtosis of the intensity distributions are measured for at least five times and their average are plotted in Figs. \ref{Figure05} (a-c). As the results clearly show the intensity distribution roughnesses associated with NV, HRS and LRS are substantially different, confirming that the samples under different voltages may contain different non-uniformities.

{

\section{Conclusion}
\label{conc}
In conclusion we introduced speckle pattern (SP) analysis for  remote, non-contact characterization of PVK memristors during their operation. 
The memristors are  fundamental devices for  neuromorphic systems and switching applications. 
We applied the technique on  ITO/PVK:rGO/Al, which  exhibits bistable resistive switching behavior. Depending on their state  they may show different resistivity in response to a constant applied voltage, and their operation can be optimized by controlling the doping \textcolor{black}{amount} of graphene oxide. The reliability and repeatability of the fabricated ITO/PVK:rGO/Al memristor were \textit{a priori} tested via examining the cyclic voltage sweep, and then they subjected to SP analysis experiments. 
The SPs which are formed through scattering of laser light from these devices, include a huge amount of information and their post-processing statistical  elaboration extracts important information about the samples, e.g.,  resistance or conduction mechanism, .
The  processing results  are expressed via several statistical parameters. The variations in the computed parameters are attributed to the  \textcolor{black}resistance state of the PVK:rGO samples under the applied voltage. The resistance mechanism  depend on the properties of PVK:rGO memristors as well as the \textcolor{black}{additional  non-uniformities induced through the variations of density of carriers}.
The present methodology has the potential to be served as a bench-top device for characterization purposes of similar samples and devices while they are in operation. 

\bibliography{2024dpoem_refs}

\section*{Acknowledgements}

The authors acknowledge the Ghazanfarian family through their financial support of the project and their donation to the Amir Alam Ghazanfarian Electronic Materials Lab at the Institute for Advanced Studies in Basic Sciences (IASBS). The authors thank Elaheh Nazari for her assistances on speckle pattern analysis.

\section*{Author contributions statement}

R.J., M.S., and B.T. carried out the experiments. R.J. analyzed and interpreted the data. D.A. and A.M. conceived and supervised the project. All authors discussed the results and contributed to the writing and reviewing of the manuscript.

\section*{Competing interests}
The authors declare no competing interests.

\section*{Data availability}
The datasets used or analyzed during the current study available from the corresponding author on reasonable request.

\end{document}